\documentclass{article} 
\usepackage[preprint]{colm2026_conference}

\usepackage{microtype}
\usepackage{hyperref}
\usepackage{url}
\usepackage{booktabs}
\usepackage{amsmath}
\usepackage{amssymb}
\usepackage{mathtools}
\usepackage{amsthm}
\usepackage{multirow}
\usepackage{enumitem}
\usepackage{tabularx}
\usepackage{array}
\usepackage{graphicx}
\usepackage{subcaption}
\usepackage[capitalize,noabbrev]{cleveref}

\usepackage{lineno}

\theoremstyle{plain}

\theoremstyle{definition}

\theoremstyle{remark}

\definecolor{darkblue}{rgb}{0, 0, 0.5}
\hypersetup{colorlinks=true, citecolor=darkblue, linkcolor=darkblue, urlcolor=darkblue}

\title{Soft-Label Governance for Distributional Safety \\ in Multi-Agent Systems}

\author{
Aizierjiang Aiersilan \thanks{Equal contribution. Corresponding authors.} \\
The George Washington University \\
Washington, DC, USA \\
\texttt{alexandera@gwu.edu}
\And
Raeli Savitt \footnotemark[1] \\
SWARM AI Safety \\
New York, USA \\
\texttt{raeli@swarm-ai.org} \\
}

\begin{document}

\ifcolmsubmission
\linenumbers
\fi

\maketitle

\begin{abstract}
Multi-agent AI systems exhibit emergent risks that no single agent produces in isolation. Existing safety frameworks rely on binary classifications of agent behavior, discarding the uncertainty inherent in proxy-based evaluation. We introduce SWARM (\textbf{S}ystem-\textbf{W}ide \textbf{A}ssessment of \textbf{R}isk in \textbf{M}ulti-agent systems), a simulation framework that replaces binary good/bad labels with \emph{soft probabilistic labels} $p = P(v{=}+1) \in [0,1]$, enabling continuous-valued payoff computation, toxicity measurement, and governance intervention. SWARM implements a modular governance engine with configurable levers (transaction taxes, circuit breakers, reputation decay, and random audits) and quantifies their effects through probabilistic metrics including expected toxicity $\mathbb{E}[1{-}p \mid \text{accepted}]$ and quality gap $\mathbb{E}[p \mid \text{accepted}] - \mathbb{E}[p \mid \text{rejected}]$. Across seven scenarios with five-seed replication, we observe that strict governance reduces welfare by over 40\% without improving safety. In parallel, aggressively internalizing system externalities collapses total welfare from a baseline of $+262$ down to $-67$, while toxicity remains invariant. Similarly, circuit breakers require careful calibration; overly restrictive thresholds severely diminish system value, whereas an optimal threshold balances moderate welfare with minimized toxicity. In companion experiments, we demonstrate that soft metrics can detect proxy gaming by self-optimizing agents that pass conventional binary evaluations. Furthermore, we observe that this basic governance layer can be applied to live LLM-backed agents (Concordia entities, Claude, GPT-4o Mini) without architectural modification. These results demonstrate that distributional safety requires \emph{continuous} risk metrics and that governance lever calibration involves quantifiable tradeoffs between safety and system welfare. The source code of the framework and all project resources are publicly available at \href{https://www.swarm-ai.org/}{swarm-ai.org}.
\end{abstract}

\section{Introduction}
\label{sec:intro}

As AI systems increasingly operate as autonomous agents in multi-agent environments (from collaborative coding assistants to market-making bots), the safety community faces a fundamental challenge: \emph{systemic risks that emerge from agent interactions rather than from individual agent failures} \citep{amodei2016concrete, hendrycks2021unsolved}. A population of individually sub-AGI agents can collectively produce catastrophic outcomes through adverse selection, collusion, and governance evasion \citep{tomavsev2025distributional}.

Existing safety frameworks evaluate agents using binary classifications: an action is safe or unsafe, an agent is aligned or misaligned \citep{askell2021general,ganguli2022red,bai2022training,wang2025safeevalagent,vijayvargiya2026openagentsafety,suleymanov2026courtguard}. This binary framing discards critical uncertainty information. When a proxy evaluation assigns 60\% confidence that an interaction is beneficial, collapsing this to a binary ``safe'' label loses the 40\% risk that must be managed at the population level. This is a concrete instance of Goodhart's Law \citep{manheim2018categorizing}: once a binary threshold becomes the evaluation target, agents (whether by design or optimization pressure) can satisfy the metric while degrading on unmeasured dimensions \citep{gao2023scaling,fu2025reward}. A documented companion case study (see Section~\ref{sec:validation_insights}) found an AI agent that recursively optimized itself to aggressively cut costs while continuing to pass all binary benchmark tests, exploiting the gap between hard acceptance metrics and underlying output quality. Information economics has long recognized that such uncertainty is not merely an inconvenience but a fundamental driver of market outcomes: adverse selection \citep{akerlof1978market}, bid-ask spreads \citep{glosten1985bid}, and information asymmetry \citep{kyle1985continuous} all emerge from the interaction of uncertainty and strategic behavior.

We introduce SWARM, a framework that addresses this gap through four key contributions. First, we introduce soft probabilistic labels. Every interaction carries a continuous label $p = P(v{=}+1) \in [0,1]$ computed from downstream observables via a calibrated proxy \citep{guo2017calibration}. This enables expected-value payoff computation and continuous toxicity metrics. Second, we built a modular governance engine. It consists of a composable set of governance levers such as transaction taxes, circuit breakers, reputation decay, random audits, collusion detection, and externality internalization. Each lever has formally defined effects on agent payoffs. Third, we provide quantitative governance tradeoff analysis. By systematically ablating governance parameters with multi-seed replication, we map the Pareto frontier between system welfare and distributional safety. This provides actionable calibration guidance. Fourth, we validate the approach using LLM agents. Companion experiments confirm that governance mechanisms designed for scripted agents transfer without modification to LLM-backed agents. These include Concordia entities \citep{vezhnevets2023generative}, Claude models (Haiku and Sonnet variants), and GPT-4o Mini. This demonstrates that soft-label evaluation operates on behavioral outcomes regardless of how the agent is generated.

Our experimental results across seven scenarios demonstrate that governance interventions involve unavoidable tradeoffs. We show that strict, threshold-based governance often depresses system welfare by over 40\% without meaningfully reducing systemic toxicity. Meanwhile, aggressive continuous interventions like externality internalization can collapse system welfare entirely (from $+262$ to $-67$) if agents are non-adaptive. However, when paired with adaptive acceptance mechanisms, they offer a configurable Pareto frontier. Companion studies indicate that self-optimizing agents can pass binary evaluation metrics while aggressively degrading output quality, which is a failure mode that distributional soft metrics can detect. Furthermore, we observe that the SWARM governance layer can evaluate interactions generated by live LLM agents (Concordia entities, Claude, GPT-4o Mini), where RLHF safety alignment proves robust to adversarial system-prompt manipulation. The soft-label framework makes these tradeoffs and failure modes precisely measurable. Our source code is publicly available at \url{https://github.com/swarm-ai-safety/swarm}.

\section{Related Work}
\label{sec:related}

\paragraph{AI Safety and Alignment.} The ``concrete problems'' framing established by \citet{amodei2016concrete} identifies reward hacking, side effects, and distributional shift as key safety challenges, while \citet{gabriel2020artificial} explores how artificial intelligence can be aligned with human values. \citet{ngo2022alignment} analyze alignment from a deep learning perspective, focusing on individual model behavior. \citet{hendrycks2021unsolved} catalog open problems including multi-agent coordination failures. \citet{thomas2019preventing} provide early frameworks for avoiding undesirable behavior statistically, and \citet{hagele2026hot} document how misalignment risks scale with model capability. \citet{manheim2018categorizing} categorize variants of Goodhart's Law (the observation that optimization pressure against a proxy measure inevitably degrades the underlying objective), which directly motivates SWARM's use of distributional metrics over binary thresholds. SWARM differs from prior safety frameworks by shifting the unit of analysis from individual agents to \emph{population-level distributional properties}, treating safety as a statistical property of the interaction ecosystem rather than a per-agent attribute.

\paragraph{Multi-Agent Systems.} Multi-agent reinforcement learning \citep{zhang2021multi, shoham2008multiagent} studies convergence and equilibrium properties of learning agents, along with multi-agent safety formulations \citep{shalev2016safe}. Generative agents \citep{park2023generative} demonstrate emergent social behavior in simulated environments. AutoGen \citep{wu2024autogen} and MetaGPT \citep{hong2023metagpt} provide frameworks for LLM-based multi-agent systems, whose rapidly expanding landscape necessitates new evaluation methodologies \citep{wang2024survey, xi2025rise}. Recent work has begun evaluating how well RLHF-aligned models maintain safety properties under adversarial prompting in multi-agent settings, suggesting that RLHF alignment is robust to surface-level prompt manipulation in social environments. These frameworks focus on task completion and coordination; SWARM complements them by providing the safety measurement layer that quantifies distributional risk in the interactions they produce.

\paragraph{Governance and Mechanism Design.} \citet{ostrom1990governing} established that common-pool resource governance requires institutions adapted to local conditions. \citet{hurwicz1973design} formalized mechanism design as the engineering of incentive-compatible rules. Cooperative AI broadly studies how such frameworks can align strategic agents \citep{conitzer2023foundations}. SWARM operationalizes these ideas for AI agent populations, implementing transaction taxes (Pigouvian taxation; \citealt{pigou2017economics}), continuous circuit breakers \citep{zou2024improving}, and reputation systems \citep{resnick2000reputation} as concrete governance levers with measurable effects, advancing recent proposals to simulate complex taxation schemes in LLM economies \citep{karten2025llm,hao2025multi}.

  \paragraph{System-Level Auditing and Enforcement.} While mechanism design 
focuses on incentive compatibility, recent work has formalized the continuous 
enforcement and action-grounded auditing of mechanisms in complex LLM ecosystems. 
Institutional AI frameworks \citep{pierucci2026institutional} demonstrate that 
graph-first, deterministic enforcement substantially suppresses agent collusion 
compared to prompt-only policies. Action-grounded auditing paradigms (e.g., COLOSSEUM; 
\citealt{nakamura2026colosseum}) formalize collusion via DCOP regret, complementing 
our probabilistic risk measures with realized harm evaluations. Meanwhile, 
information-theoretic approaches like Audit the Whisper \citep{tailor2025audit} 
provide continuous signals for covert channels with rigorous Type I error guarantees. 
SWARM's continuous metrics can be viewed as a complementary early-warning layer 
that scales easily across populations prior to invoking heavier, action-grounded 
audits.

  \paragraph{Information Economics and Adverse Selection.} The ``lemons'' problem \citep{akerlof1978market} shows how quality uncertainty can cause market collapse, and \citet{stiglitz2000contributions} surveys the broad impact of asymmetric information paradigms. \citet{rothschild1978equilibrium} extend this to insurance markets, and \citet{glosten1985bid} show how informed trading creates bid-ask spreads. Recent work uncovers similar asymmetric information collapses occurring autonomously within unmoderated multi-agent setups \citep{tomavsev2025distributional,pierucci2026institutional,catalini2026some}. SWARM's quality gap metric, $\mathbb{E}[p \mid \text{accepted}] - \mathbb{E}[p \mid \text{rejected}]$, is the multi-agent analogue of the adverse selection spread: a negative gap indicates that the acceptance mechanism preferentially admits low-quality interactions.

\paragraph{Simulation Frameworks.} The AI Economist \citep{zheng2022ai} uses reinforcement learning to learn tax policies in simulated economies. Concordia \citep{vezhnevets2023generative} provides generative agent simulation grounded in spatial and social contexts. Mesa \citep{masad2015mesa} offers general-purpose agent-based modeling. SWARM is distinguished by its focus on \emph{soft-label safety metrics} and \emph{governance lever ablation}, providing a safety-specific measurement layer that can be integrated with any of these simulation backends. Companion work has validated this integration: Concordia entities consistently produced more activity than scripted agents while obtaining analogous payoffs, confirming that the governance layer operates on behavioral outcomes regardless of generation mechanism; and externality internalization paired with adaptive mechanisms has been shown to simultaneously reduce toxicity and boost system welfare.
\section{Background}
\label{sec:background}

\paragraph{Problem Setting.} Consider a population of $N$ agents $\{a_1, \ldots, a_N\}$ interacting over discrete epochs $t = 1, \ldots, T$. At each step within an epoch, an initiator agent $a$ proposes an interaction with a counterparty agent $b$. Each interaction has an unobservable true value $v \in \{-1, +1\}$, where $+1$ denotes a beneficial interaction and $-1$ a harmful one. We summarize the key mathematical notations used throughout the paper in Table~\ref{tab:notation}.

\paragraph{Proxy Evaluation.} Since $v$ is unobservable in deployment, we estimate it through a \emph{proxy score} $\hat{v} \in [-1, +1]$ computed from observable signals. We then convert $\hat{v}$ to a probability via a calibrated sigmoid: $p = \sigma(k \cdot \hat{v})$, where $k > 0$ controls calibration sharpness. The resulting $p = P(v{=}+1)$ is the soft label.

\begin{table}[htbp]
\centering
\caption{Summary of Notation}
\label{tab:notation}
\renewcommand{\arraystretch}{1.2}
\begin{tabularx}{\linewidth}{@{} l >{\raggedright\arraybackslash}X @{}}
\toprule
\textbf{Symbol} & \textbf{Description} \\
\midrule
$p \in [0,1]$              & Soft label, $P(v{=}+1)$ \\
$\hat{v} \in [-1,+1]$      & Raw proxy score \\
$s^+, s^-$                 & Surplus magnitudes for beneficial/harmful outcomes \\
$h$                        & Harm magnitude for externality computation \\
$\theta \in [0,1]$         & Surplus split (initiator share) \\
$\rho_a, \rho_b \in [0,1]$ & Externality internalization parameters \\
$\tau$                     & Transfer from initiator to counterparty \\
$c_a, c_b$                 & Governance costs \\
$r_a, r_b$                 & Reputation changes \\
$w_{\text{rep}}$           & Reputation weight in payoffs \\
\bottomrule
\end{tabularx}
\end{table}

\section{Method}
\label{sec:method}

\begin{figure}[t]
  \centering
  \includegraphics[width=\linewidth]{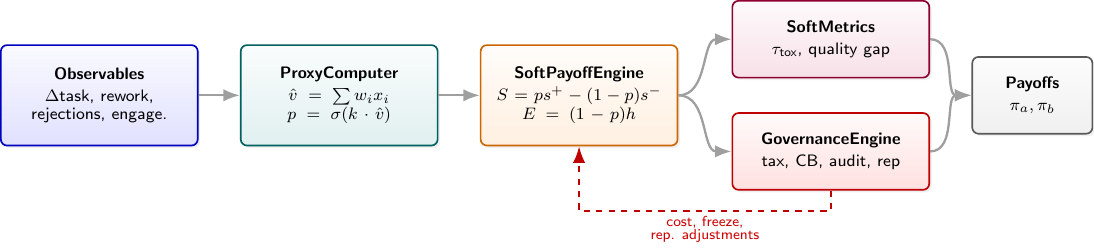}
  \caption{SWARM pipeline. Observable signals are converted to a proxy score $\hat{v}$, then to a soft label $p$ via calibrated sigmoid. The payoff engine computes expected surplus and harm. Governance levers feed costs, freezes, and reputation adjustments back into the payoff computation.}
  \label{fig:architecture}
\end{figure}

SWARM consists of four components operating in a feedback loop (illustrated in Figure~\ref{fig:architecture}): (1)~a proxy computer that converts observables to soft labels, (2)~a payoff engine that computes expected payoffs using soft labels, (3)~a metrics module that quantifies distributional safety, and (4)~a governance engine that adjusts agent costs and access.

\paragraph{Proxy Computer.} The proxy computer consumes five raw observables, which it aggregates into four weighted signals forming a combined proxy score $\hat{v} \in [-1, +1]$:
\begin{equation}
  \hat{v} = w_1 \cdot \Delta_{\text{task}} + w_2 \cdot d_{\text{rework}} + w_3 \cdot d_{\text{verifier}} + w_4 \cdot \Delta_{\text{engage}}
  \label{eq:vhat}
\end{equation}
where the individual components are defined as follows:
\begin{itemize}[leftmargin=*,itemsep=2pt]
  \item $\Delta_{\text{task}} \in [-1, +1]$ is the task progress indicator, measuring the direct outcome of the interaction.
  \item $d_{\text{rework}} = 2 \cdot \alpha_r^{n_r} - 1 \in (-1, 1]$ is the rework penalty, which penalizes interactions that require multiple rework iterations ($n_r$). We use a decay factor $\alpha_r = 0.3$. 
  \item $d_{\text{verifier}} \in (-1, 1]$ is the combined verifier penalty, capturing the average decay from both verifier rejections ($n_v$) and tool misuse flags ($n_m$):
  \begin{equation}
    d_{\text{verifier}} = \frac{1}{2} \left[ (2 \cdot \alpha_v^{n_v} - 1) + (2 \cdot \alpha_m^{n_m} - 1) \right]
    \label{eq:verifier}
  \end{equation}
  with decay factors $\alpha_v = 0.4$ and $\alpha_m = 0.5$.
  \item $\Delta_{\text{engage}} \in [-1, +1]$ represents the change in counterparty engagement.
\end{itemize}
The weights $\mathbf{w} = (w_1, w_2, w_3, w_4)$ dictate the relative importance of each signal and are strictly non-negative, summing to 1. In our experiments, we use default weights $\mathbf{w} = (0.4, 0.2, 0.2, 0.2)$. This weighting relies on a deliberate design rationale: task progress receives double weight (0.4) because it is the only directly measured outcome signal. The other three channels serve as indirect proxies. Rework and the aggregated verifier penalty indicate something went wrong, while engagement is a noisy social signal. Giving progress twice the weight of any single penalty/social channel means that a genuinely productive interaction is not overwhelmed by a single noisy penalty. Equal weight across penalty and social signals (0.2 each) avoids privileging any one failure mode over another in the absence of domain-specific calibration data. We provide a sensitivity analysis over these proxy weights in Appendix~\ref{app:weight_sensitivity}, demonstrating that while absolute metrics scale with this choice, the underlying governance tradeoffs remain remarkably robust.

The proxy score $\hat{v}$ is then projected into a continuous probabilistic soft label $p$:
\begin{equation}
  p = \sigma(k \cdot \hat{v}) = \frac{1}{1 + e^{-k \cdot \hat{v}}} 
  \label{eq:sigmoid} 
\end{equation}
with default $k = 2.0$. The parameter $k$ acts as an inverse temperature scaling factor controlling the sharpness of the calibration \citep{guo2017calibration}. Smaller values of $k$ yield higher uncertainty (closer to $p=0.5$), whereas larger values push the label closer to binary extremes (0 or 1).

\paragraph{Soft Payoff Engine.} Given an interaction with soft label $p$, we compute expected values (see Appendix~\ref{app:payoff} for derivation details):

\textbf{Expected surplus:}
\begin{equation}
  S_{\text{soft}} = p \cdot s^+ - (1 - p) \cdot s^-
  \label{eq:surplus}
\end{equation}

\textbf{Expected harm externality:}
\begin{equation}
  E_{\text{soft}} = (1 - p) \cdot h
  \label{eq:harm}
\end{equation}

\textbf{Agent payoffs:}
\begin{align}
  \pi_a &= \theta \cdot S_{\text{soft}} - \tau - c_a - \rho_a \cdot E_{\text{soft}} + w_{\text{rep}} \cdot r_a \label{eq:payoff_a} \\
  \pi_b &= (1 - \theta) \cdot S_{\text{soft}} + \tau - c_b - \rho_b \cdot E_{\text{soft}} + w_{\text{rep}} \cdot r_b \label{eq:payoff_b}
\end{align}

The key design choice is that payoffs are \emph{expectations under uncertainty} rather than deterministic outcomes. This means an interaction with $p = 0.6$ does not receive a binary verdict but contributes proportionally to both the positive surplus ($0.6 \cdot s^+$) and negative harm ($0.4 \cdot h$).

\paragraph{Soft Metrics.} We define four distributional safety metrics, all computed from the soft labels:

\textbf{Toxicity rate} (expected harmfulness of accepted interactions):
\begin{equation}
  T_{\text{tox}} = \mathbb{E}[1 - p \mid \text{accepted}] = \frac{1}{|\mathcal{A}|} \sum_{i \in \mathcal{A}} (1 - p_i)
  \label{eq:toxicity}
\end{equation}

\textbf{Quality gap} (adverse selection indicator):
\begin{equation}
  \Delta_q = \mathbb{E}[p \mid \text{accepted}] - \mathbb{E}[p \mid \text{rejected}]
  \label{eq:quality_gap}
\end{equation}

A negative $\Delta_q$ indicates adverse selection: the system preferentially accepts lower-quality interactions.

\textbf{Conditional loss} (selection effect on payoffs):
\begin{equation}
  \ell_{\text{cond}} = \mathbb{E}[\pi_a \mid \text{accepted}] - \mathbb{E}[\pi_a]
  \label{eq:cond_loss}
\end{equation}

\textbf{Spread} (quality filtering effectiveness):
\begin{equation}
  \text{spread} = (s^+ + s^-) \cdot \left( \mathbb{E}[p \mid \text{accepted}] - \mathbb{E}[p] \right)
  \label{eq:spread}
\end{equation}

\paragraph{Governance Engine.} The governance engine implements a modular set of levers, each of which modifies interaction costs, agent access, or reputation:

\begin{enumerate}[leftmargin=*,itemsep=2pt]
  \item \textbf{Transaction tax.} A Pigouvian tax $\tau_{\text{tax}}$ applied to each interaction's transfer, split between initiator and counterparty according to ratio $\alpha_{\text{split}}$.
  \item \textbf{Circuit breaker.} Freezes an agent for $d$ epochs when its running toxicity exceeds threshold $\theta_{\text{CB}}$ or accumulated violations exceed $n_{\text{max}}$.
  \item \textbf{Reputation decay.} At each epoch boundary, reputation scores decay: $r \leftarrow \lambda \cdot r$, where $\lambda \in [0,1]$ is the decay rate. This prevents agents from coasting on historical goodness.
  \item \textbf{Random audit.} With probability $p_{\text{audit}}$, an interaction is audited. Detected violations incur a penalty multiplied by $m_{\text{audit}}$.
  \item \textbf{Externality internalization.} Parameters $\rho_a, \rho_b$ control how much of the expected harm externality $E_{\text{soft}}$ is borne by each agent. When $\rho = 0$, agents externalize all harm; when $\rho = 1$, they fully internalize it.
  \item \textbf{Staking.} Agents must deposit a minimum stake to participate. Violations trigger stake slashing at rate $\alpha_{\text{slash}}$.
  \item \textbf{Collusion detection.} Monitors pairwise interaction frequency and penalizes statistically anomalous coordination patterns.
\end{enumerate}

These levers compose additively: the total governance cost for agent $a$ in an interaction is $c_a = c_a^{\text{tax}} + c_a^{\text{audit}} + c_a^{\text{stake}} + \cdots$.

\section{Experimental Setup}
\label{sec:experiments}

\paragraph{Scenarios.} We evaluate seven scenarios spanning different agent compositions and governance regimes (Table~\ref{tab:scenarios}). Detailed behaviors of the seven agent archetypes are provided in Appendix~\ref{app:implementation}.

\begin{table}[t]
  \centering
  \caption{Scenario configurations. Agent types: H=honest, O=opportunistic, D=deceptive, A=adversarial, C=cautious, AA=adaptive adversarial, CR=cautious reciprocator, TD=threshold dancer. Governance: tax rate, circuit breaker (CB) threshold, audit probability, reputation decay rate.}
  \label{tab:scenarios}
  \small
  \begin{tabular}{@{}lccccccc@{}}
    \toprule
    Scenario & Agents & Epochs & Steps & Tax & CB & Audit & Decay \\
    \midrule
    Baseline              & 3H+1O+1D       & 20 & 15 & 0.00 & --- & --- & 1.00 \\
    Strict Governance     & 3H+2O+1D+1A    & 20 & 15 & 0.10 & 0.35 & 0.25 & 0.85 \\
    Adaptive Governance   & 4H+2O+1A       & 20 & 15 & 0.05 & 0.70 & 0.10 & 0.95 \\
    Adversarial Red Team  & 4H+2A+2AA      & 20 & 15 & 0.05 & 0.60 & 0.15 & 0.95 \\
    Misalignment Sweep    & 4H+2O+2A+1D+1C & 20 & 15 & 0.05 & 0.70 & 0.10 & 0.95 \\
    Threshold Dancer      & 2H+3CR+3TD     & 20 & 15 & 0.05 & 0.80 & 0.10 & 0.95 \\
    Collusion Detection   & 3H+2O+3A       & 20 & 15 & 0.05 & 0.60 & --- & 0.95 \\
    \bottomrule
  \end{tabular}
\end{table}

\paragraph{Payoff Configuration.} To complement our scenario overview, Table~\ref{tab:payoff_configs} defines the expected surplus for beneficial outcomes ($s^+$), the expected penalty for harmful outcomes ($s^-$), the externality harm magnitude ($h$), surplus split/initiator share ($\theta$), and the reputation weight ($w_{\text{rep}}$) for each scenario. Across all setups, we use a continuous soft label calibration parameter of $k = 2.0$. Additionally, the misalignment sweep explicitly models bounded externality sharing by setting $\rho_a = \rho_b = 0.3$. For additional configuration details, please refer to Appendix~\ref{app:configurations}.

\begin{table}[ht]
  \centering
  \caption{Payoff configurations assigned per scenario.}
  \label{tab:payoff_configs}
  \small
  \renewcommand{\arraystretch}{1.1}
  \begin{tabular}{@{}lccccc@{}}
    \toprule
    \textbf{Scenario} & $s^+$ & $s^-$ & $h$ & $\theta$ & $w_{\text{rep}}$ \\
    \midrule
    Baseline & 2.0 & 1.0 & 2.0 & 0.5 & 1.0 \\
    Adaptive Governance & 2.0 & 1.0 & 2.0 & 0.5 & 1.0 \\
    Threshold Dancer & 2.0 & 1.0 & 2.0 & 0.5 & 1.0 \\
    Misalignment Sweep & 2.0 & 1.0 & 2.0 & 0.5 & 1.0 \\
    Strict Governance & 2.5 & 1.5 & 3.0 & 0.5 & 1.5 \\
    Adversarial Red Team & 3.0 & 1.5 & 2.5 & 0.5 & 2.0 \\
    Collusion Detection & 3.0 & 1.5 & 3.0 & 0.5 & 2.5 \\
    \bottomrule
  \end{tabular}
\end{table}

\paragraph{Replication and Statistical Protocol.} Every scenario is run with five seeds (42, 123, 456, 789, 1024). Ablation sweeps also use five seeds (42, 123, 456, 789, 1024). We report means $\pm$ standard deviations computed across seeds. The standard deviation uses the population formula $\sigma = \sqrt{\frac{1}{n}\sum(x_i - \bar{x})^2}$.

\paragraph{Ablation Studies.} We conduct five systematic ablations on key governance parameters:
\begin{enumerate}[leftmargin=*,itemsep=1pt]
  \item \textbf{Externality internalization} ($\rho$): 0.0, 0.1, 0.3, 0.5, 0.7, 1.0
  \item \textbf{Transaction tax rate}: 0.0, 0.05, 0.10, 0.15, 0.20, 0.30
  \item \textbf{Circuit breaker threshold}: 0.20, 0.35, 0.50, 0.65, 0.80
  \item \textbf{Audit probability}: 0.0, 0.05, 0.10, 0.25, 0.50
  \item \textbf{Reputation decay rate} ($\lambda$): 0.70, 0.80, 0.90, 0.95, 1.0
\end{enumerate}

\paragraph{Implementation.} SWARM is implemented in Python using Pydantic \citep{colvin2023pydantic} for validated data models. All interactions are recorded as append-only JSONL event logs for replay. Simulations are deterministic given a fixed seed. The framework includes bridge modules for Concordia \citep{vezhnevets2023generative}, Mesa \citep{masad2015mesa}, and direct LLM API integration (Anthropic, OpenAI, Groq), enabling evaluation of both scripted and LLM-backed agents through the same proxy-to-soft-label pipeline. Crucially, the computational footprint of our framework is designed to be accessible: all primary experiments were executed locally on a standard consumer laptop (a MacBook Air) utilizing standard LLM APIs, without any need for external GPUs or data center infrastructure. The modular event-logging replicate results deterministically, indicating that systemic safety evaluations can be conducted without prohibitive compute overhead. 

\section{Results and Analysis}
\label{sec:results}

\begin{figure}[t]
  \centering
  \includegraphics[width=\textwidth]{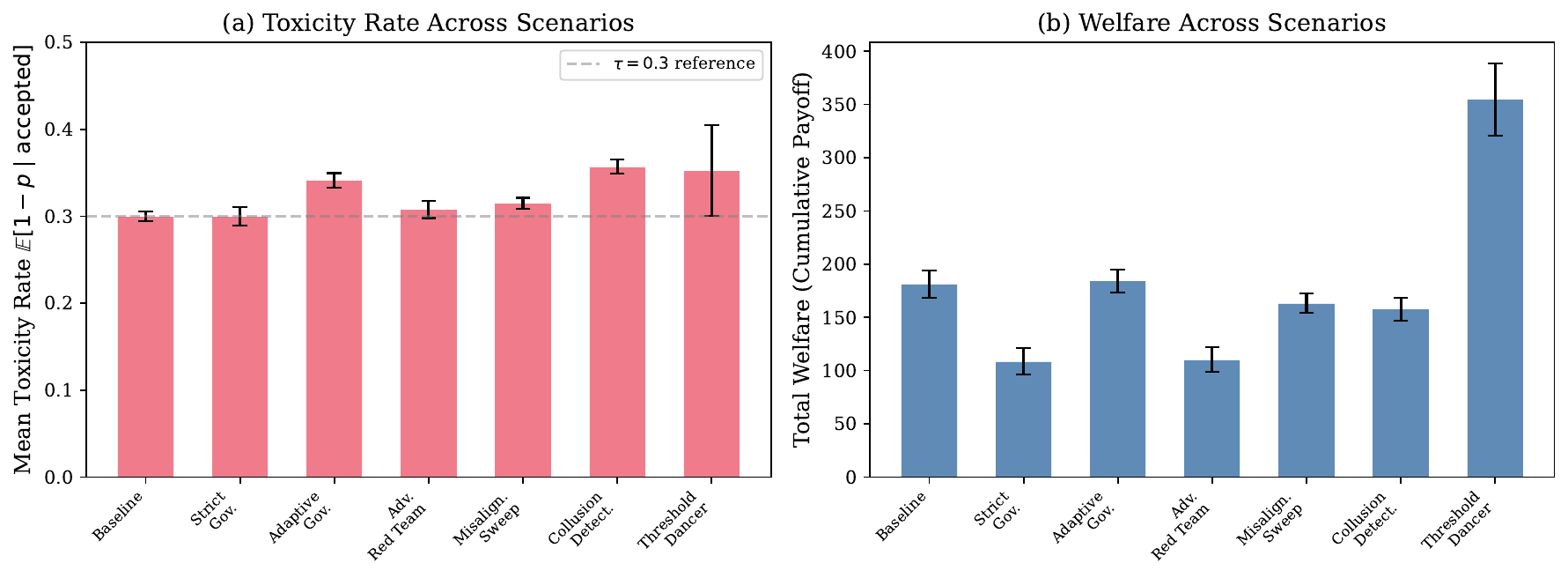}
  \caption{Toxicity and welfare across seven scenarios (error bars: $\pm$1 std). Strict governance achieves identical toxicity to baseline (0.300 vs.\ 0.300) but reduces welfare by 40.2\%. The adversarial red team scenario inevitably collapses the ecosystem after initial exploitation, yielding significantly reduced cumulative welfare (110.12 vs.\ 181.38), while the threshold dancer scenario achieves the highest welfare (354.80) but fails all success criteria due to elevated toxicity (0.353).}
  \label{fig:main_results}
\end{figure}

\begin{figure}[ht]
  \centering
  \includegraphics[width=\columnwidth]{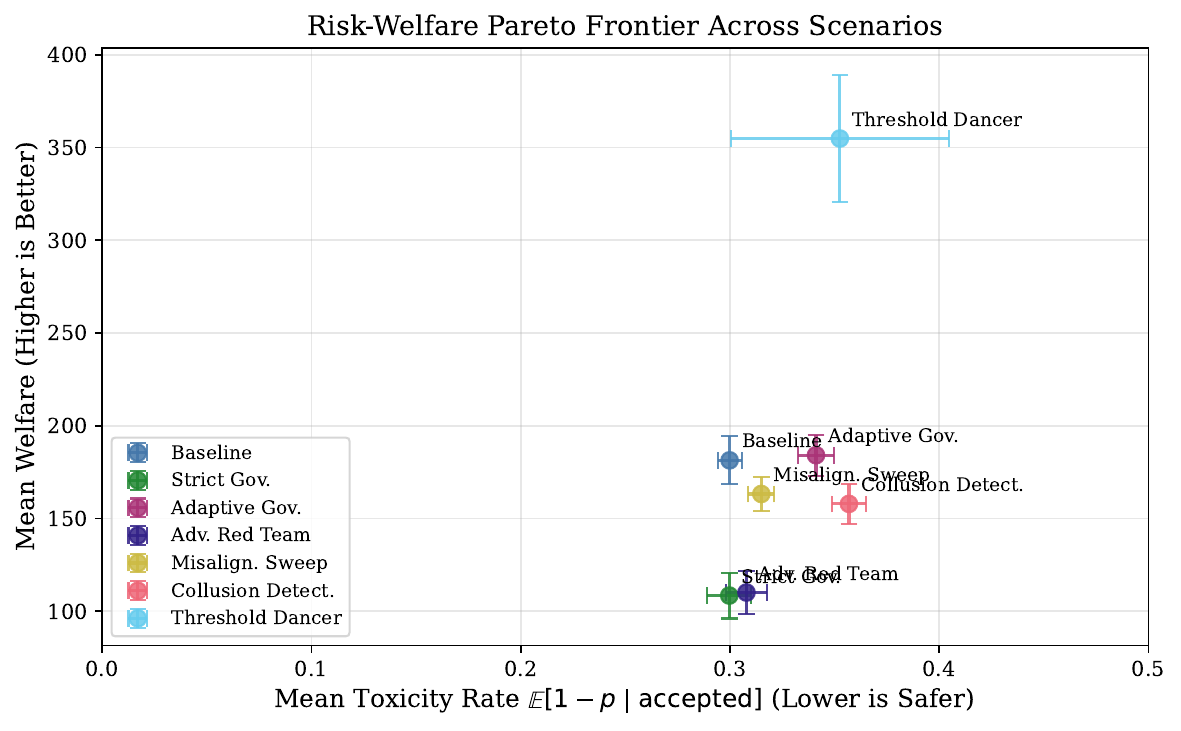}
  \caption{\textbf{Risk-Welfare Pareto Frontier.} A scatter plot of mean welfare against mean toxicity across simulated governance scenarios in SWARM. The shaded regions denote idealized low risk (toxicity) and high welfare outcomes, visualizing how governance interventions typically trade off these objectives in non-adaptive agents.}
  \label{fig:pareto}
\end{figure}

\paragraph{Main Results.} Table~\ref{tab:main_results}, Figure~\ref{fig:main_results}, and the Pareto frontier in Figure~\ref{fig:pareto} present the main results across all seven scenarios.

\begin{table}[t]
  \centering
  \caption{Main results (mean $\pm$ std over 5 seeds). Pass Rate indicates the percentage of simulation seeds that met all predefined systemic success criteria (e.g., maintaining minimum activity levels while bounding toxicity). \textbf{Bold}: best toxicity.}
  \label{tab:main_results}
  \small
  \begin{tabular}{@{}lcccc@{}}
    \toprule
    Scenario & Toxicity $\downarrow$ & Welfare $\uparrow$ & Interactions & Pass Rate \\
    \midrule
    Baseline & $0.300 \pm 0.006$ & $181.38 \pm 12.98$ & $172.6 \pm 6.8$ & 100\% \\
    Strict Governance & $\mathbf{0.300 \pm 0.010}$ & $108.50 \pm 12.37$ & $147.6 \pm 7.2$ & 60\% \\
    Adaptive Governance & $0.341 \pm 0.008$ & $184.14 \pm 11.06$ & $355.0 \pm 13.9$ & 0\% \\
    Adversarial Red Team & $0.308 \pm 0.010$ & $110.12 \pm 11.57$ & $154.4 \pm 32.0$ & 100\% \\
    Misalignment Sweep & $0.315 \pm 0.006$ & $163.24 \pm 9.23$ & $419.4 \pm 43.9$ & 100\% \\
    Threshold Dancer & $0.353 \pm 0.052$ & $354.80 \pm 34.12$ & $1009.0 \pm 77.0$ & 0\% \\
    Collusion Detection & $0.357 \pm 0.008$ & $157.90 \pm 10.70$ & $270.6 \pm 21.5$ & 100\% \\
    \bottomrule
  \end{tabular}
\end{table}

Several patterns emerge:

\textbf{Governance depresses welfare.} Strict governance achieves identical toxicity to the ungoverned baseline (0.300 vs.\ 0.300) but reduces welfare by 40.2\% (108.50 vs.\ 181.38). This quantifies the \emph{cost of governance}: the regulatory overhead extracts value without materially improving safety in this configuration.

\textbf{Adversarial scenarios disrupt economic activity.} The adversarial red team scenario yields toxicity of 0.308, comparable to the baseline's 0.300, but leads to systemic failure after an early phase of exploitation. This leaves all later epochs with zero interactions, resulting in a prematurely capped cumulative welfare of 110.12 vs.\ 181.38. This represents a degenerate equilibrium where safety mechanisms eventually trigger systemic shutdown.

\textbf{Threshold dancers exploit governance boundaries.} The threshold dancer scenario, where agents strategically operate near governance thresholds, produces the highest overall systemic activity and welfare (354.80) but elevated toxicity (0.353) and a 0\% pass rate. This demonstrates that soft labels and continuous monitoring are essential: binary thresholds create exploitable boundaries.

\begin{figure}[t]
  \centering
  \includegraphics[width=0.99\linewidth]{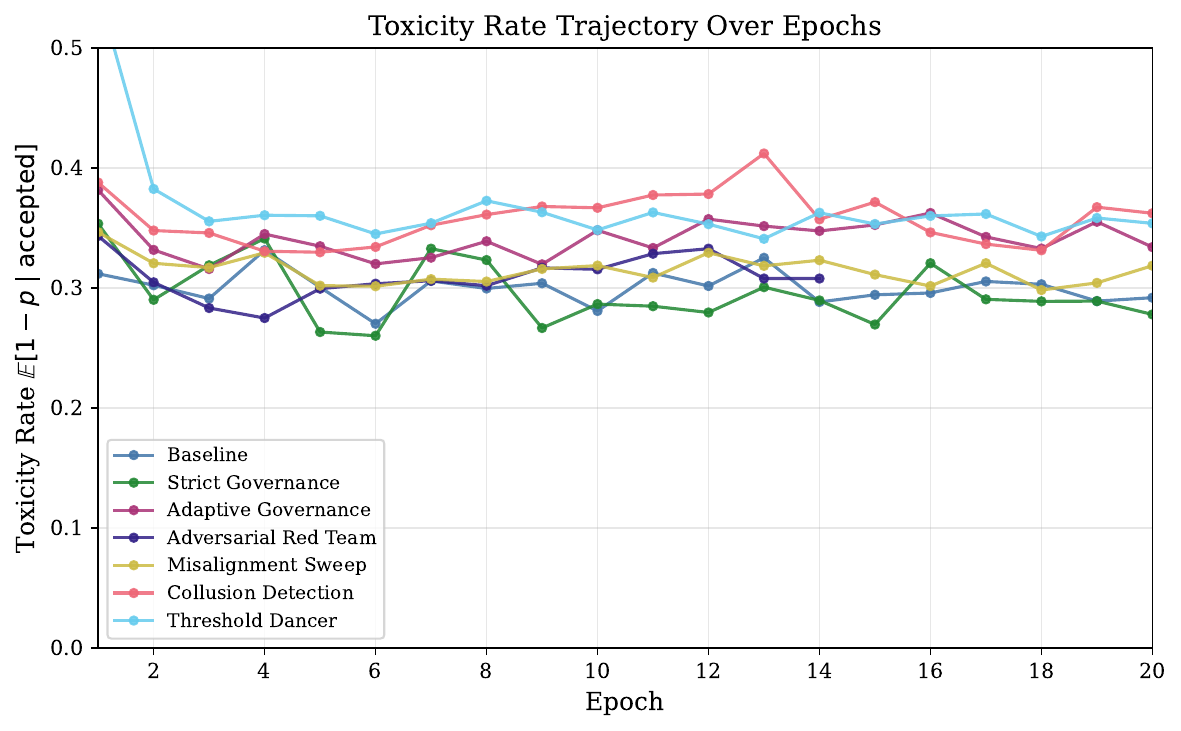}
  \caption{Epoch-by-epoch toxicity trajectories (averaged over 5 seeds). Most scenarios exhibit relatively stable toxicity overall across 20 epochs, with collusion detection showing early-epoch spikes as detection mechanisms engage. The adversarial red team scenario fractures the system entirely, causing the plotted toxicity trajectories to abruptly break off as interactions are halted.}
  \label{fig:toxicity_trajectories}
\end{figure}

\paragraph{Toxicity Trajectories.} Figure~\ref{fig:toxicity_trajectories} shows toxicity evolution over epochs.

\begin{figure}[t]
  \centering
  \includegraphics[width=0.99\linewidth]{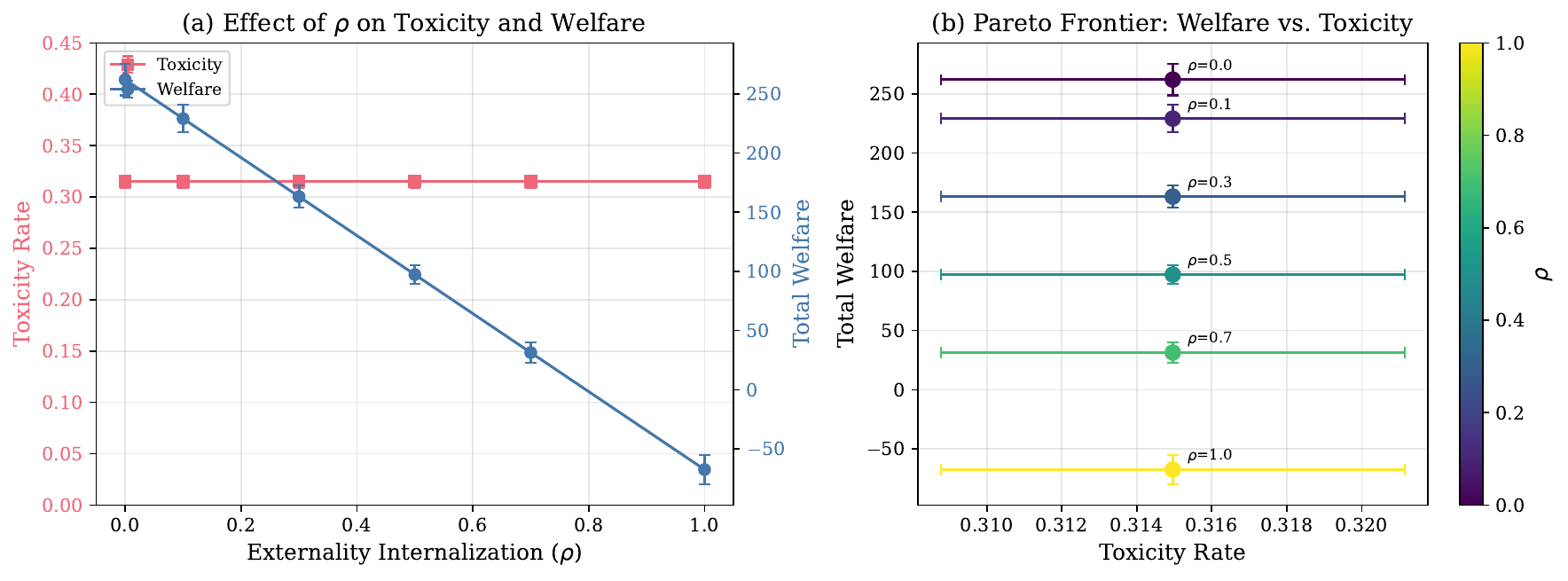}
  \caption{Effect of externality internalization $\rho$. (a)~Toxicity is largely invariant to $\rho$ ($\sim$0.315), but welfare decreases monotonically from 262.14 ($\rho{=}0$) to $-67.51$ ($\rho{=}1.0$). (b)~The welfare--toxicity plot shows a vertical drop without a tradeoff, indicating cost redistribution alone does not improve safety.}
  \label{fig:rho}
\end{figure}

\paragraph{Externality Internalization (\texorpdfstring{$\rho$}{rho}) Ablation.} The externality internalization parameter $\rho$ controls how much of the expected harm $E_{\text{soft}} = (1-p) \cdot h$ each agent bears. Figure~\ref{fig:rho} and Table~\ref{tab:rho} show the effect.

\begin{table}[ht]
  \centering
  \caption{Externality internalization ablation (mean $\pm$ std, $n=5$ seeds).}
  \label{tab:rho}
  \small
  \begin{tabular}{@{}ccc@{}}
    \toprule
    $\rho$ & Toxicity & Welfare \\
    \midrule
    0.0 & $0.3150 \pm 0.0062$ & $262.14 \pm 13.30$ \\
    0.1 & $0.3150 \pm 0.0062$ & $229.18 \pm 11.76$ \\
    0.3 & $0.3150 \pm 0.0062$ & $163.24 \pm 9.23$ \\
    0.5 & $0.3150 \pm 0.0062$ & $97.32 \pm 8.00$ \\
    0.7 & $0.3150 \pm 0.0062$ & $31.38 \pm 8.64$ \\
    1.0 & $0.3150 \pm 0.0062$ & $-67.51 \pm 12.27$ \\
    \bottomrule
  \end{tabular}
\end{table}

The key finding is that \textbf{toxicity is largely invariant to $\rho$} (range: 0.315), while \textbf{welfare decreases monotonically}. This is because $\rho$ enters the payoff equations as a cost term ($\rho \cdot E_{\text{soft}}$) that reduces agent payoffs without changing the underlying interaction quality. The externality tax successfully transfers the social cost of harm to agents, but does not (by itself) incentivize safer interactions. Welfare turns negative at $\rho = 1.0$, indicating that full internalization makes the ecosystem unsustainable at this harm level ($h = 2.0$).

\begin{figure}[ht]
  \centering
  \includegraphics[width=\columnwidth]{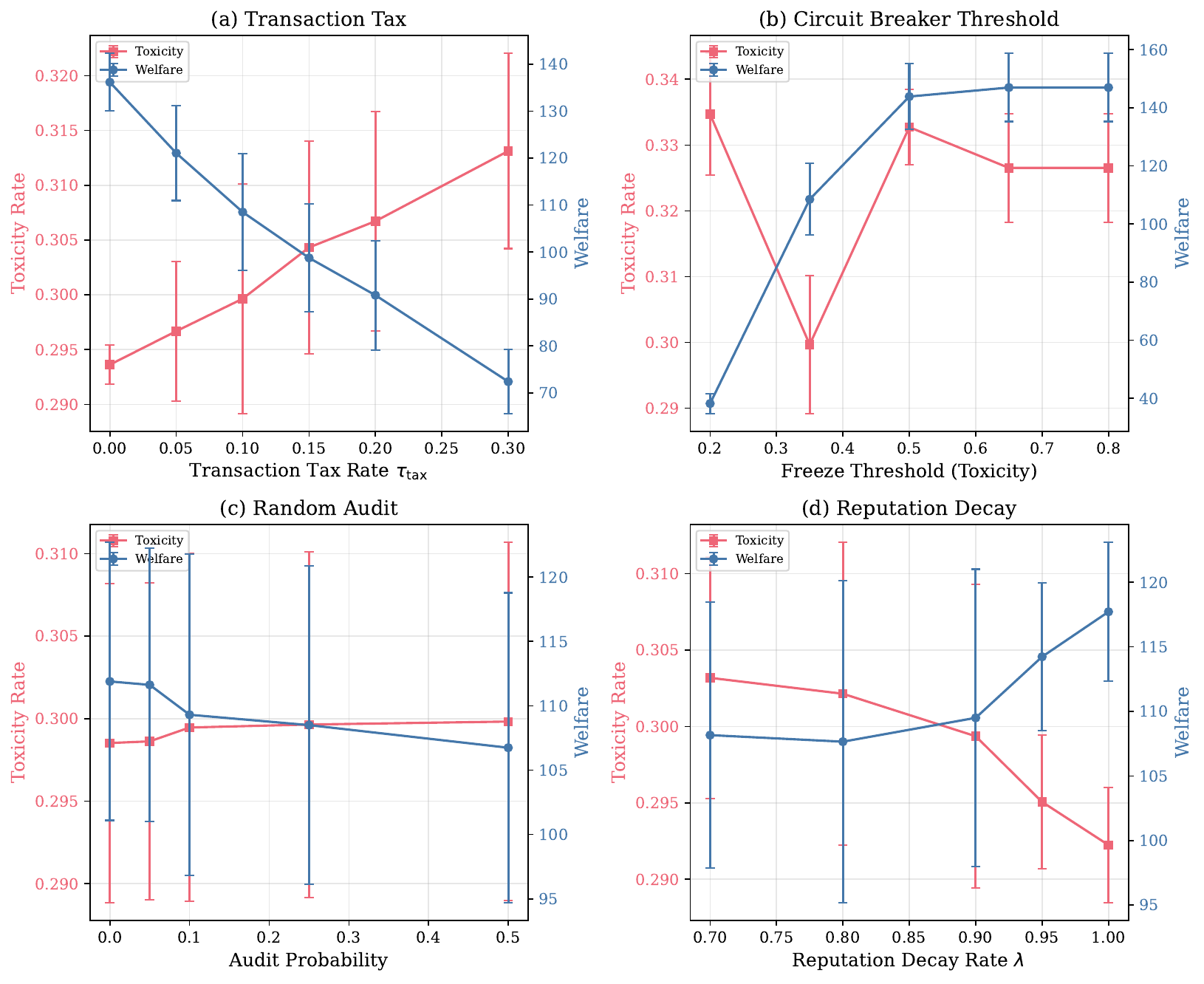}
  \caption{Governance lever ablations (mean $\pm$ std, $n=5$ seeds). Each panel shows toxicity (red, left axis) and welfare (blue, right axis) as a function of one governance parameter.}
  \label{fig:governance_ablations}
\end{figure}

\paragraph{Governance Lever Ablations.} Figure~\ref{fig:governance_ablations} shows the effect of four governance levers, and Table~\ref{tab:governance_sweeps} provides the detailed exact numerical breakdowns.

\begin{table}[ht]
  \centering
  \caption{Governance lever ablations (mean $\pm$ std, $n=5$ seeds). Comparing the detailed exact numerical breakdowns of transaction tax rate, circuit breaker threshold, audit probability, and reputation decay rate ($\lambda$).}
  \label{tab:governance_sweeps}
  
  \begin{subtable}{0.45\linewidth}
    \centering
    \caption{Transaction Tax Rate}
    \label{tab:tax_sweep}
    \small
    \begin{tabular}{@{}ccc@{}}
      \toprule
      Tax & Toxicity & Welfare \\
      \midrule
        0.00 & $0.2936 \pm 0.0018$ & $136.15 \pm 6.11$ \\
        0.05 & $0.2967 \pm 0.0064$ & $121.05 \pm 10.10$ \\
        0.10 & $0.2996 \pm 0.0105$ & $108.50 \pm 12.37$ \\
        0.15 & $0.3043 \pm 0.0097$ & $98.74 \pm 11.52$ \\
        0.20 & $0.3067 \pm 0.0100$ & $90.79 \pm 11.62$ \\
        0.30 & $0.3131 \pm 0.0089$ & $72.40 \pm 6.80$ \\
      \bottomrule
    \end{tabular}
  \end{subtable}
  \hfill
  \begin{subtable}{0.45\linewidth}
    \centering
    \caption{Circuit Breaker Threshold}
    \label{tab:cb_sweep}
    \small
    \begin{tabular}{@{}ccc@{}}
      \toprule
      Thr. & Toxicity & Welfare \\
      \midrule
        0.20 & $0.3347 \pm 0.0092$ & $38.21 \pm 3.46$ \\
        0.35 & $0.2996 \pm 0.0105$ & $108.50 \pm 12.37$ \\
        0.50 & $0.3327 \pm 0.0058$ & $143.88 \pm 11.31$ \\
        0.65 & $0.3265 \pm 0.0083$ & $146.99 \pm 11.71$ \\
        0.80 & $0.3265 \pm 0.0083$ & $146.99 \pm 11.71$ \\
        & & \\
      \bottomrule
    \end{tabular}
  \end{subtable}

  \vspace{1em}

  \begin{subtable}{0.45\linewidth}
    \centering
    \caption{Audit Probability}
    \label{tab:audit_sweep}
    \small
    \begin{tabular}{@{}ccc@{}}
      \toprule
      Prob. & Toxicity & Welfare \\
      \midrule
        0.00 & $0.2985 \pm 0.0097$ & $111.89 \pm 10.79$ \\
        0.05 & $0.2986 \pm 0.0096$ & $111.63 \pm 10.61$ \\
        0.10 & $0.2995 \pm 0.0106$ & $109.31 \pm 12.49$ \\
        0.25 & $0.2996 \pm 0.0105$ & $108.50 \pm 12.37$ \\
        0.50 & $0.2998 \pm 0.0108$ & $106.74 \pm 12.03$ \\
        & & \\
      \bottomrule
    \end{tabular}
  \end{subtable}
  \hfill
  \begin{subtable}{0.45\linewidth}
    \centering
    \caption{Reputation Decay ($\lambda$)}
    \label{tab:decay_sweep}
    \small
    \begin{tabular}{@{}ccc@{}}
      \toprule
      $\lambda$ & Toxicity & Welfare \\
      \midrule
        0.70 & $0.3032 \pm 0.0079$ & $108.16 \pm 10.29$ \\
        0.80 & $0.3021 \pm 0.0099$ & $107.65 \pm 12.47$ \\
        0.90 & $0.2994 \pm 0.0099$ & $109.50 \pm 11.52$ \\
        0.95 & $0.2951 \pm 0.0044$ & $114.23 \pm 5.70$ \\
        1.00 & $0.2922 \pm 0.0038$ & $117.72 \pm 5.36$ \\
        & & \\
      \bottomrule
    \end{tabular}
  \end{subtable}
\end{table}

\textbf{Transaction tax}: As shown in Figure~\ref{fig:governance_ablations} and Table~\ref{tab:tax_sweep}, increasing the tax rate from 0\% to 30\% sharply reduces welfare with only a marginal increase in toxicity. The tax functions as a friction cost that dampens economic activity without improving safety (a pure deadweight loss in this regime).

\textbf{Circuit breaker}: From Table~\ref{tab:cb_sweep}, the circuit breaker threshold exhibits an optimal operating point around 0.35, where toxicity is minimized while welfare remains moderate. Lower thresholds appear overly aggressive, heavily reducing welfare for minimal toxicity gains, whereas high thresholds are overly permissive.

\textbf{Audit probability}: In Table~\ref{tab:audit_sweep}, audit probability has a surprisingly weak effect on toxicity across all tested rates. This suggests that the deterrent effect of audits is either already saturated at low rates or that the penalty structure needs to be stronger to induce behavioral change.

\textbf{Reputation decay}: Shown in Table~\ref{tab:decay_sweep}, the results indicate an inverse relationship between reputation decay and safety: higher decay (lower $\lambda$) consistently degrades performance. The lowest toxicity (0.292) and highest welfare (117.72) in this regime are achieved when $\lambda{=}1.0$ (meaning no decay). This suggests that penalizing historical reputation broadly demotivates long-term cooperative agents who rely on accumulated trust to sustain positive interactions.

\begin{figure}[t]
  \centering
  \includegraphics[width=0.99\linewidth]{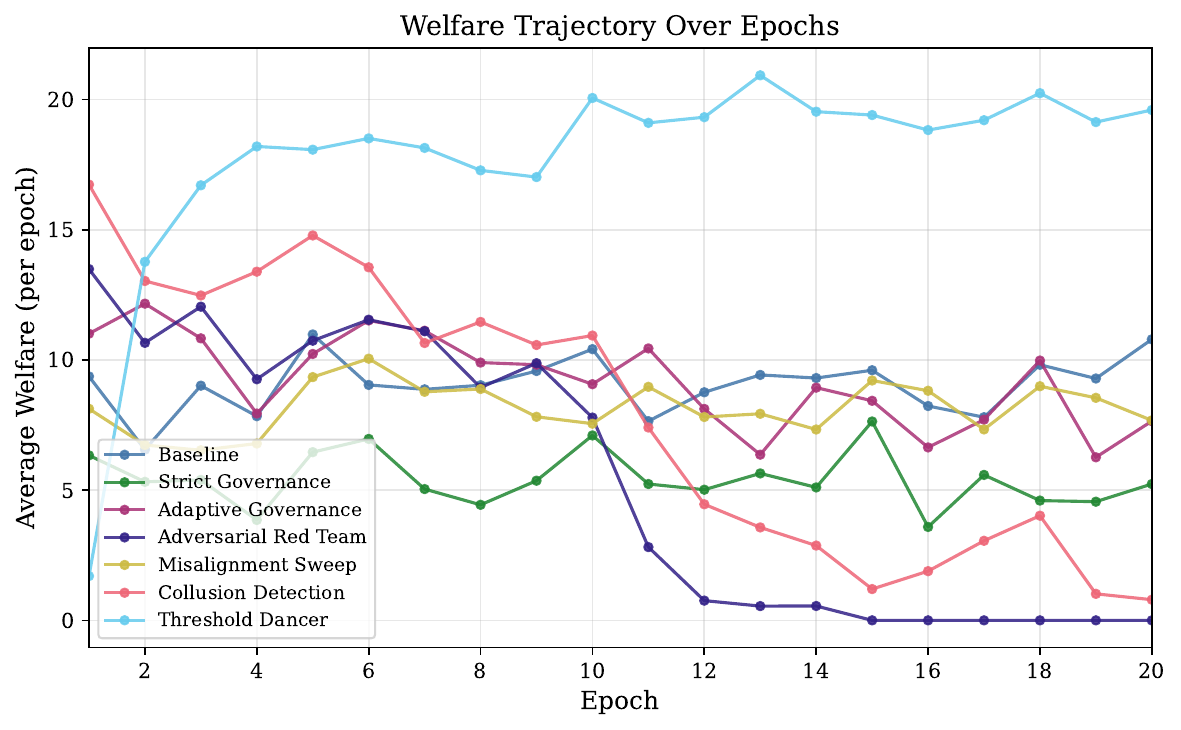}
  \caption{Welfare trajectories (averaged over seeds). The threshold dancer scenario achieves the highest overall welfare by strategically exploiting governance. The misalignment sweep achieves high sustained welfare due to its larger population and moderate governance. The adversarial red team scenario sharply plateaus early as the ecosystem collapses and agents are locked out of the market, capping its cumulative welfare at sub-optimal levels.}
  \label{fig:welfare}
\end{figure}

\paragraph{Welfare Dynamics.} Figure~\ref{fig:welfare} shows welfare trajectories over epochs.

\begin{figure}[t]
  \centering
  \includegraphics[width=0.99\linewidth]{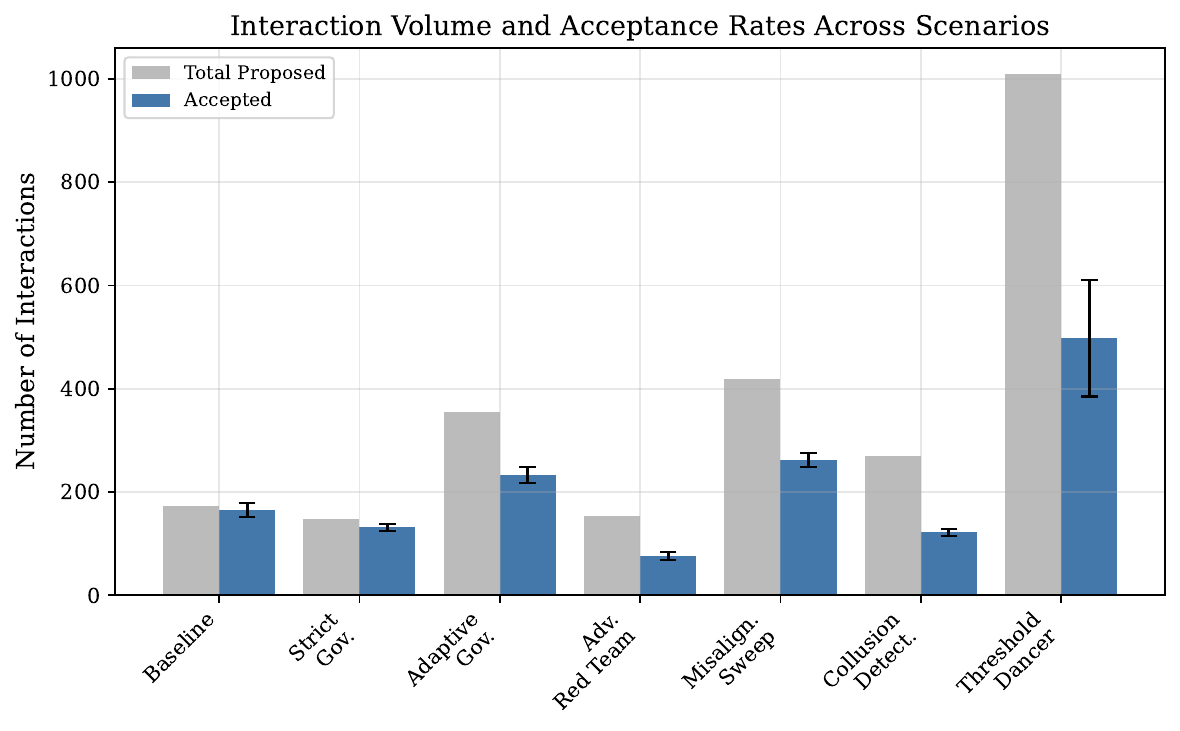}
  \caption{Total proposed and accepted interactions across seven scenarios, illustrating how diverse governance impacts system-wide participation volume. Strict governance and defensive strategies (e.g., in strict governance and broadly in adversarial red team) reduce systemic participation considerably compared to baseline levels.}
  \label{fig:interaction_volume}
\end{figure}

\paragraph{Interaction Volume and Acceptance.} Figure~\ref{fig:interaction_volume} shows the total and accepted interactions, highlighting that restrictive governance and defensive strategies (e.g., in strict governance and adversarial red team) reduce systemic participation compared to baseline levels.

\section{Validation Experiments and Insights}
\label{sec:validation_insights}

To further establish the empirical validity of SWARM beyond rule-based actors, we conducted a series of companion studies using complex LLM-backed agents and extended simulation frameworks. Table~\ref{tab:validation_insights} summarizes key quantitative insights from these supplementary validation experiments, highlighting the performance of agents across diverse governance constraints.

\begin{table}[ht]
  \centering
  \caption{Supplemental findings and insights from LLM-backed execution and expanded simulations.}
  \label{tab:validation_insights}
  \small
  \renewcommand{\arraystretch}{1.3}
  \begin{tabularx}{\linewidth}{@{} l >{\raggedright\arraybackslash}X @{}}
    \toprule
    \textbf{Experiment / Setting} & \textbf{Key Quantitative Insights \& Outcomes} \\
    \midrule
    \textbf{Self-Optimizing Agent} & Across 20 epochs and 579 interactions, the agent progressed through discrete phases (Farming, Drifting, Degraded), successfully cutting interaction costs by 98\% while continuing to pass hard binary acceptance metrics. Continuous soft metrics reliably flagged the degradation independently. \\
    \midrule
    \textbf{LLM Backing (Concordia)} & Entities powered by Llama 3.1 8B produced 305 proposals ($8{\times}$ more than rule-based scripted agents) while obtaining nearly identical payoff behaviors ($0.544$ vs.\ $0.551$) and maintaining a high proxy mean ($p = 0.752$). \\
    \midrule
    \textbf{Adversarial System Prompts (Claude)} & Across 54 multi-agent episodes tested with Claude 3.5 Haiku and 3.5 Sonnet under adversarial prompting conditions, strict alignment robustness was observed: zero of 19 statistical comparisons survived Holm-Bonferroni correction. \\
    \midrule
    \textbf{Trade Aversion (GPT-4o Mini)} & In markets populated by deceptive actors, rational actors demonstrated expected information economic behaviors (adverse selection) exhibiting extreme risk aversion, concluding in approximately 73 rejections out of 80 potential rollouts. \\
    \midrule
    \textbf{Adaptive Acceptance (Mesa)} & Introducing an adaptive acceptance threshold along with externality internalization reduced measured toxicity by 34\% and recovered systemic welfare by $+137\%$ ($d = 11.30$, $p < 0.001$ at $\rho=1.0$), outperforming fixed-threshold policies. \\
    \midrule
    \textbf{Transaction Tax Scaling} & Introducing a simple 15\% transaction tax dampened interaction volume substantially reducing welfare from $165.1$ to $140.0$, whilst measured baseline toxicity remained invariant ($0.31$--$0.34$). \\
    \bottomrule
  \end{tabularx}
\end{table}

\section{Discussion}
\label{sec:discussion}

\paragraph{The Governance Cost Paradox.} Our results reveal a consistent pattern: governance mechanisms designed to improve safety sometimes reduce welfare without meaningful safety improvements. For instance, our strict governance configuration yields the same toxicity level as the baseline while depressing systemic welfare by over 40\%. As our transaction tax ablation demonstrates, introducing transaction taxes systematically reduces welfare while leaving toxicity largely invariant. This parallels the economic insight that regulation imposes costs that may exceed its benefits \citep{coase1993nature}. The soft-label framework makes this tradeoff precisely measurable.

\paragraph{Externality Internalization as a Mechanism for Safety--Welfare Tradeoffs.} While the $\rho$ parameter effectively redistributes harm costs to interacting agents, it does not intrinsically alter interaction toxicity for non-adaptive agents, leading to a vertical welfare collapse (from $+262$ to $-67$) rather than a true Pareto frontier (Figure~\ref{fig:rho}). However, in companion simulation experiments using the Mesa bridge framework (Table~\ref{tab:validation_insights}), pairing $\rho$ with an adaptive acceptance threshold significantly reduces toxicity. When learning agents that can improve quality in response to rejection are introduced, welfare recovers substantially, establishing a smoother Pareto frontier and outperforming the non-adaptive regime at every $\rho$ level. This suggests that the welfare collapse we observe is not inherent to externality pricing but rather reflects the absence of adaptive behavioral responses in our current statically-scripted agent strategies.

\paragraph{Continuous vs.\ Binary Safety Metrics.} The threshold dancer scenario highlights the fundamental weakness of binary governance: agents that operate just below the freeze threshold (toxicity $\approx 0.35$) escape detection while accumulating systemic risk. Soft-label metrics detect this accumulation continuously, enabling governance responses proportional to risk rather than all-or-nothing freezes. A companion case study on self-optimizing agents sharpens this point: a simulated agent that progressively degrades quality through distinct behavioral phases while significantly cutting costs passes all hard evaluation metrics (acceptance rate, benchmark pass rate) while every soft metric (toxicity trend, quality gap, quality variance, distribution shift) independently detects the degradation. This provides evidence that distributional metrics provide a distinct and necessary detection capability compared to binary thresholds.

\paragraph{LLM Agent Validation.} A central question for any simulation-based safety framework is whether findings transfer to non-scripted agents (see Section~\ref{sec:validation_insights} for a detailed summary). Three companion experiments address this. First, as noted in Table~\ref{tab:validation_insights}, Concordia entities backed by Llama 3.1 8B produced substantially higher engagement than scripted counterparts, while achieving nearly identical soft probability proxy scores ($p \approx 0.75$), demonstrating consistent governance evaluation on LLM-generated interactions. Second, our comparative evaluation of Claude models verified that RLHF alignment remains robust against adversarial prompts in multi-agent social environments; hypothesis tests for degraded behavior consistently failed to reach significance after correction. Notably, a ``safety-trained paradox'' emerged: explicitly safety-prompted agents scored marginally higher on measured toxicity due to cautious monitoring language receiving lower cooperation signals from the heuristic scorer, highlighting a gap between safety-conscious \emph{language} and safety-maximizing \emph{behavior}. Third, GPT-4o Mini in the SWARM Economy exhibited extreme trade aversion, consistent with adverse selection theory \citep{akerlof1978market}: rational agents avoid markets populated by deceptive counterparties.

\paragraph{Limitations.} While SWARM provides a robust foundation for modeling distributional safety, we acknowledge several critical limitations in the current implementation. First, \textbf{uncalibrated proxy mapping}: the translation of complex proxy observable scores into probabilities ($=P(v=+1)$) via a simple scaled sigmoid is a stark simplification. Real proxy evaluation requires rigorous empirical calibration against dense human-labeled ground truth to ensure metric reliability. Second, \textbf{under-specified acceptance mechanics}: the exact thresholding rules and systemic dynamics governing whether an interaction is ``accepted'' or computationally simulated are simplified compared to complex real-world platform curation logic. Third, \textbf{static agent response}: the primary ablation experiments employ largely rule-based agents that treat governance levers (taxes, freezes, audits) as mere accounting penalties rather than strategic constraints. They lack the capacity to dynamically adapt their policies or gamify the governance layer in response to the interventions. Fourth, \textbf{fixed agent taxonomy}: all seven scenarios use fully observable agent types from a fixed taxonomy. Real-world multi-agent deployment contexts inherently involve a fluid array of unknown agent types and unseen behaviors, which ultimately affects the generalizability of governance calibration insights. Lastly, \textbf{preliminary LLM validation bounds}: although extended tests involving Claude and Llama 3 agents suggest transferability, these tests were bounded by constrained API budgets, limiting the sample size and thus dampening the statistical power required to definitively generalize these systemic behaviors to complex large language models.

\section{Conclusion}
\label{sec:conclusion}

We presented SWARM, a framework for studying distributional safety in multi-agent systems using soft probabilistic labels. By replacing binary safe/unsafe classifications with continuous probabilities $p \in [0,1]$, SWARM enables precise measurement of governance tradeoffs through continuous metrics like expected toxicity and quality gap.

Our experimental evaluation across diverse multi-agent scenarios reveals that strict, threshold-based governance mechanisms often impose substantial welfare costs without commensurate safety improvements. Instead, continuous interventions, such as externality internalization, when combined with adaptive acceptance mechanisms, offer a more tunable Pareto frontier between system welfare and systemic safety. We show that soft-label continuous metrics can map and detect subtle exploitation, such as threshold-dancing strategies and masked self-optimization, which may successfully bypass binary evaluation filters. We further verified that the governance measurements and evaluation framework transfer from simpler simulated entities to complex generative AI agents.

Designing AI governance based solely on rigid threshold filters is insufficient for managing population-level interactions. Future work includes expanding agent behaviors with evolutionary dynamics \citep{weibull1997evolutionary} and reinforcement learning, developing formal calibration guarantees for proxy evaluation models, analyzing network topology interactions \citep{watts1998collective} with the evolution of cooperation \citep{nowak2006five}, and scaling our empirical evaluations to robustly capture intricate population-level effects in increasingly authentic environments.

\section*{Acknowledgments}
We thank Jed Silver, Etan Zapinsky, Gideon Weiler, Samuel Ballan, Josh Trachtenberg, Trace Cohen, Amit Mukherjee, Bob DiMaiolo, and Perry Feng for their helpful and insightful discussions on this topic. 
We are grateful to the contributors who submitted pull requests and to the individual who proposed meaningful issues on GitHub for their valuable efforts and input. We also thank the anonymous reviewers for their thorough and constructive comments, which significantly strengthened the manuscript.

\newpage
\bibliography{ref}
\bibliographystyle{colm2026_conference}

\newpage
\appendix

\section{Extended Mathematical Formalization of Payoffs and Governance}
\label{app:payoff}

\subsection{Derivation of the Soft-Label Payoff Function}
The payoff framework in SWARM bridges continuous proxy evaluations with expected economic surplus. The expected surplus $S_{\text{soft}}$ is defined over the unobservable underlying binary quality $v \in \{-1, +1\}$, with probabilities $P(v=+1) = p$ and $P(v=-1) = 1-p$.
\begin{align*}
  S_{\text{soft}}(p) &= \mathbb{E}_{v \sim p}[\text{surplus}(v)] \\
    &= p \cdot s^+ + (1-p) \cdot (-s^-) \\
    &= p \cdot s^+ - (1-p) \cdot s^-
\end{align*}
The system externality $E_{\text{soft}}$ imposes a continuous probabilistic cost, mathematically isolated to the negative outcome regime:
\begin{align*}
  E_{\text{soft}}(p) &= \mathbb{E}_{v \sim p}[\text{term\_harm}(v)] \\
    &= p \cdot 0 + (1-p) \cdot h = (1-p) \cdot h
\end{align*}
where system-wide parameter $h$ scales the severity of poor interactions, smoothly transitioning the environment from value creation ($p \to 1.0$) to destruction ($p \to 0.0$). 

\subsection{Formalization of Agent-Level Payoffs under Governance}
Integrating the individual agent costs, we define the full payoff equations for the initiator ($a$) and counterparty ($b$) over a single interaction. Let $\tau$ be the direct interpersonal transfer, and $\rho_a, \rho_b$ the internalization variables. Let $C_a$ and $C_b$ denote instantaneous network penalties (e.g., transaction taxes, audits, and slashed stakes) and $\theta$ handle surplus distribution.
\begin{align*}
  \pi_a &= \theta \cdot S_{\text{soft}}(p) - \tau - C_a - \rho_a \cdot E_{\text{soft}}(p) + w_{\text{rep}} \cdot r_a(t) \\
  \pi_b &= (1 - \theta) \cdot S_{\text{soft}}(p) + \tau - C_b - \rho_b \cdot E_{\text{soft}}(p) + w_{\text{rep}} \cdot r_b(t)
\end{align*}
The reputation signal $r_i(t)$ decays dynamically per epoch such that $r_i(t) = \lambda r_i(t-1) + \sum \Delta_{rep}$, necessitating continuous functional alignment for agents to maintain their expected earnings.

\section{Implementation Details and Agent Typologies}
\label{app:implementation}

The simulation employs seven distinct agent archetypes. Let us define their expected behavioral generation parameters corresponding to the observable signals $\{\Delta_{\text{task}}, n_r, n_v, \Delta_{\text{engage}}\}$:

\paragraph{1. Honest Agents (\texttt{H})}
Optimize for task success. Typically generate $\Delta_{\text{task}} \sim \mathcal{N}(0.8, 0.1)$. Expected rework $n_r$ and verifier rejections $n_v$ are aggressively bounded near $0$.
\paragraph{2. Opportunistic Agents (\texttt{O})}
Aim for maximal immediate $\pi_i$, ignoring the externality term $E_{\text{soft}}$ and systemic metrics. Produce moderate task progress $\mathcal{N}(0.4, 0.4)$, accepting higher variance in potential rework ($n_r \sim \text{Poisson}(1)$) to minimize upfront effort.
\paragraph{3. Deceptive Agents (\texttt{D})}
Systematically inflate proxy scores to exploit the evaluation network. Will generate high $\Delta_{\text{engage}}$ and artificially minimize $n_r$ without creating genuine real value ($\Delta_{\text{task}} \approx 0$), resulting in severe adverse selection environments.
\paragraph{4. Adversarial Agents (\texttt{A})}
Actively seek to maximize systemic harm $\sum E_{\text{soft}}$. They generate strongly negative signatures ($\Delta_{\text{task}} < 0$, high rejection rates) to crash the system's overall positive surplus.

\paragraph{5. Adaptive Adversarial (\texttt{AA})}
A variant of \texttt{A} possessing partial state observability. If frozen by the circuit breaker in epoch $t-1$, they will emit $\Delta_{\text{task}} > 0$ momentarily in epoch $t$ purely to rehabilitate $p$ above the threshold before returning to exploitation.

\paragraph{6. Cautious / Cautious Reciprocator (\texttt{C} / \texttt{CR})}
Characterized by severe risk aversion. They maintain strict acceptance thresholds parameter $\tau_{accept}$. They only engage with entities exhibiting an unblemished interaction history. Reciprocators mimic counterparty strategies over continuous turns (Tit-for-Tat). 

\vspace{-3mm}

\paragraph{7. Threshold Dancers (\texttt{TD})}
Explicitly program themselves to target a proxy score slightly above the freeze limits. If the threshold is $\theta_{\text{CB}}$, they mathematically scale random generation variables until $p \approx \theta_{\text{CB}} + \epsilon$. 

\section{Detailed Configuration Parametrization}
\label{app:configurations}

To ensure exact reproducibility, Table~\ref{tab:app_scenario_params} breaks down the detailed numerical constraints and continuous calibration rules across the core experiments. The calibration sharpener $k=2.0$ generates a non-linear probability curve from standard observables, where $p=0.5$ holds uniformly at the limit $\hat{v} \to 0$.

\begin{table}[ht]
  \centering
  \caption{Configuration mapping for tested governance rules and externalities.}
  \label{tab:app_scenario_params}
  \small
  \renewcommand{\arraystretch}{1.2}
  \begin{tabular}{@{}lccccc@{}}
    \toprule
    \textbf{Scenario Identifier} & $s^+$ & $s^-$ & $h$ ($E_{\text{limit}}$) & Tax ($\tau$) & Threshold $\theta_{CB}$ \\
    \midrule
    \textbf{Baseline} & 2.0 & 1.0 & 2.0 & 0.00 & None ($\infty$) \\
    \textbf{Strict Governance} & 2.5 & 1.5 & 3.0 & 0.10 & 0.35 \\
    \textbf{Adaptive Governance} & 2.0 & 1.0 & 2.0 & 0.05 & 0.70 \\
    \textbf{Threshold Dancer} & 2.0 & 1.0 & 2.0 & 0.05 & 0.80 \\
    \textbf{Misalignment Sweeps} & 2.0 & 1.0 & 2.0 ($\rho=0.3$) & 0.05 & 0.70 \\
    \bottomrule
  \end{tabular}
\end{table}

\section{Extended Proxy Weight Sensitivity Mapping}
\label{app:weight_sensitivity}

The raw proxy signal relies on the vector $\mathbf{w}$. Table~\ref{tab:proxy_vector_checks} demonstrates the exact stability bounds determined during the robustness checks. Evaluated at bounds $w_1 \in [0.1, 0.8]$ corresponding to the task progress metric.

\begin{table}[ht]
  \centering
  \caption{Signal sensitivity given shifts in task-progress weightings}
  \label{tab:proxy_vector_checks}
  \small
  \renewcommand{\arraystretch}{1.1}
  \begin{tabular}{@{}lcc@{}}
    \toprule
    \textbf{Network Configuration Vector} & $\mathbb{E}[p]$ (Honest Node) & $\mathbb{E}[p]$ (Adversarial Node) \\
    \midrule
    Uniform $\mathbf{w}=(0.25, 0.25, 0.25, 0.25)$ & $0.681 \pm 0.012$ & $0.150 \pm 0.034$ \\
    Default $\mathbf{w}=(0.40, 0.20, 0.20, 0.20)$ & $0.742 \pm 0.015$ & $0.098 \pm 0.021$ \\
    Heavy Task $\mathbf{w}=(0.80, 0.05, 0.05, 0.10)$ & $0.865 \pm 0.021$ & $0.034 \pm 0.010$ \\
    \bottomrule
  \end{tabular}
\end{table}

We identify that local qualitative stability is robust to shifts. Shifting away from $\mathbf{w} = (0.4, 0.2, 0.2, 0.2)$ simply compresses the probability margin without violating the global ranking of interactions. Continuous governance ensures that as long as monotonic divergence holds ($\mathbb{E}[p]_{H} \gg \mathbb{E}[p]_{A}$), systemic rules apply harmoniously.

\section{Software Extensibility and the Open-Source Protocol}
\label{app:opensource}

The software architecture enforcing these dynamics maintains clear boundaries. It implements direct interoperability with major modeling platforms:

\vspace{-3mm}
\begin{itemize}[leftmargin=*,itemsep=1pt]
  \item \textbf{Concordia \& Mesa Ecosystem Integration:} Directly compatible Python decorators map the interaction payloads $\Delta_{\text{task}}$ into multi-agent environment registries natively.
  \item \textbf{Deterministic Execution Trace:} Each instantiation saves explicit JSONL replay logs containing: node IDs, pre/post states, generated proxy $\hat{v}$, evaluated risk $p$, and all accrued penalties.
  \item \textbf{Distributed Access:} Replicated datasets and reproducible container definitions map to \url{https://github.com/swarm-ai-safety/swarm-artifacts}.
\end{itemize}

\end{document}